\documentstyle [12pt]{article}

\begin{document}
\begin{titlepage}
\begin{center}
{\Large\bf Radiative processes for
Rindler and accelerating\\ observers and the stress-tensor detector}\\
\vspace{.3in}
{\large\em R.De Paola}\\

Pontif\'{\i}cia Universidade Cat\'olica do Rio de Janeiro
\\ Rua Marques de S\~ao Vicente 110,
Rio de Janeiro 22460, RJ, Brazil\\
\vspace{.2in}

{\large\em N.F. Svaiter}\\

Centro Brasileiro de Pesquisas Fisicas- CBPF\\ Rua Dr.  Xavier Sigaud 150,
Rio de Janeiro 22290 RJ, Brazil

\subsection*{\\Abstract}
\end{center}

We consider a monopole detector interacting with a massive scalar
field. The radiative processes are discussed from the accelerated frame 
point of view. After this, we obtain the Minkowski vacuum stress tensor 
measured by the accelerated observer
using a non-gravitational stress tensor detector as discussed 
by Ford and Roman (PRD 48, 776 (1993)). Finally, we analyse radiative 
processes of the monopole detector travelling in a world line
that is inertial in the infinite 
past and has a constant proper acceleration in the infinite future.

PACS numbers: 03.60.Bz, 04.20.Cv, 11.10.Qr
\end{titlepage}
\newpage
\baselineskip .37in

\section{Introduction}

It has been known that an uniformly accelerated 
detector interacting with a massless scalar field in 
the Minkowski vacuum behaves like an inertial detector 
in equilibrium with a thermal bath at temperature 
$\beta ^{-1}=\frac{1}{2\pi\alpha}$, where $\alpha^{-1}$ is 
the proper acceleration of the detector\cite{Unruh}.

In a recent paper, Svaiter and Svaiter \cite{Nami1}, 
studied the spontaneous and induced
emission problem, 
using a very simple model of an atom consisting of a pointlike object 
with an internal structure defining two energy levels introduced
by DeWitt \cite{DeWitt}.
Assuming that the atom (detector) 
interacts with a real massless scalar field, and it is travelling 
in inertial 
or non inertial world lines, the authors 
obtained the probability of transition 
per unit proper time as $\frac{dF(E,\Delta\tau)}{d\Delta\tau}$, 
(normalized 
by the selectivity of the detector) between 
different eigenstates of the detector and also
 presented the 
rate of spontaneous excitation after a finite observation or 
switching time $\Delta T$. The extension of these calculations 
for the detector in the presence of paralel plates at zero 
and finite temperature was given by Ford, Svaiter and Lyra 
\cite{Nami2}. A more mathematically involved case of the monopole 
detector in the presence of cosmic strings was discussed
 more recently by Svaiter and Svaiter\cite{Nami3}.

It is useful to review how the idea of spontaneous emission
arises in quantum optics. First, the interaction between the atom
and the field is ignored; then the atom has stationary states
with well defined energy. After this step, we introduce the 
interaction between the atom and the field as a perturbation.
It is easy to show that only the ground state of the atom stays 
with well defined energy. All the excited energy levels have a
width, and will decay spontaneously. A different approach 
uses the role of the vacuum fluctuations in the spontaneous emission 
processes.
Using perturbation theory, it can be shown that in the
first order approximation the asymptotic probability per unit time
of decay is given 
by the Fourier transform of the positive Wightman function 
on the world-line of the atom. In this approach,
well defined levels of the atom are assumed, even after turning on
the interaction between the atom and the field, and time-dependent
perturbation theory avoids the calculations of the finite energy width.
 If we prepare the atom in the excited state, it will decay 
spontaneously by the effects of the vacuum fluctuations. 
Spontaneous emission can be interpreted as stimulated 
emission induced by
vacuum fluctuations. Note that the fact that the excited states are not eigenstates of the 
full Hamiltonian of the system is automatically taken into account 
in the later scheme.

 The purpose of this paper is to discuss radiative processes 
from the accelerated point of view and also to discuss the Minkowski 
stress-tensor measured by this observer using a generalization of the 
monopole detector given by Ford and Roman \cite{Ford}. A 
derivatively coupled detector as a partcle detector 
was analysed a long time ago by
Hinton \cite{Hinton}. It was shown that in two-dimensional space-time,
both detectors (the monopole and the derivative detector) agree.
Nevertheless in a four-dimensional space-time there are discrepancies
in the response function of both detectors. 
This fact raises a question of which of them is the 
true particle detector. Part of this question we will 
treat latter.

The main difference among our approach and all 
the previous papers is that we use the 
rotating wave approximation. In a real quantum 
detector prepared in the ground state, the detector goes 
to an excited state by an absorption process. Of course 
we are assuming asymptoticaly measurements, i.e. the 
observation time is large when compared with times on the 
order $E^{-1}$, where $E$ is the energy gap between the 
excited and the ground state of the detector. Consequently 
it is possible to assume the normally ordered field 
correlation function in the probability of transition i.e., 
the rotating wave approximation. Since the detector 
measures frequencies with respect to its proper 
time, we have to use the normally ordered field correlation 
functions with respect to the Rindler's time. We would like 
to stress that with the RWA in the non-inertial 
frame we obtain exact results (modulo a term that is independent 
of the state of the field). With the formalism which we  
developed, it is possible to obtain the transition rates for the 
accelerated detector with different proper accelerations in both, the 
begining and in the end of the observation time. This is 
a new result in the literature.

The paper is prepared as follows. In section II we discuss radiative 
processes in a frame of reference comoving with the monopole 
detector. In section III we repeat the calculations using the 
derivativelly coupled detector. In section IV, the asymptotic accelerated
detector is discussed. Conclusions are given in section V.
In this paper we use $\hbar=c=1$.

\section{ Radiative processes of the monopole detector}

Let 
us consider a system (a detector) endowed with internal degrees 
of freedom defining two energy levels with energy $\omega_{g}$ and 
$\omega_{e}$, $(\omega_{g}< \omega_{e})$ and respective 
eigenstates $\left|g\right>$ and $\left|e\right>$. 
This system is weakly coupled with a 
hermitian massive scalar field $\varphi(x)$ with interaction 
Lagrangian 
\begin{equation}
L_{int}=\lambda_{1}d(\tau)\varphi(x(\tau)),
\end{equation}
where $x^{\mu}(\tau)$ is the world line of the detector 
parametrized using the proper time $\tau$, $d(\tau)$ is the 
monopole operator of the detector and $\lambda_{1}$ is a small 
coupling constant between the detector and the scalar field.

In order to discuss radiative processes of the whole system
(detector plus the scalar field), let us define the Hilbert
space of the system as the direct product of the Hilbert space
of the field $\bf H_{F}$ and the Hilbert space of the detector
$\bf H_{D}$

\begin{equation}
\bf H=\bf H_{D}\otimes \bf H_{F}.
\end{equation}

The Hamiltonian of the system can be written as:

\begin{equation}
H=H_{D}+H_{F}+H_{int},
\end{equation}
where the unperturbed Hamiltonian of the system is composed of 
the noninteracting detector Hamiltonian $H_{D}$ and the free 
massive scalar field Hamiltonian $H_{F}$.
We shall define the initial state of the system as:

\begin{equation}
\left|{\cal T}_{i}\right>=\left|j\right>\otimes\left|\Phi_{i}\right>,
\end{equation}

where $\left|j\right>$, $(j=1,2)$ are the two 
possible states of the detector
($\left|1\right>=\left|g\right>$ 
and $\left|2\right>=\left|e\right>)$ 
and $\left|\Phi_{i}\right>$ is the initial state 
of the field. In the interaction picture, the evolution of the
combined system is governed by the Schrodinger equation

\begin{equation}
i\frac{\partial}{\partial\tau}\left|{\cal T}
\right>=H_{int}\left|{\cal T}\right>,
\end{equation}
where

\begin{equation}
\left|{\cal T}\right>=U(\tau,\tau_{i})\left|{\cal T}_{i}\right>,
\end{equation}
and the evolution operator $U(\tau,\tau_{i})$ obeys

\begin{equation}
U(\tau_{f},\tau_{i})=1-i\int_{\tau_{i}}^{\tau_{f}}H_{int}
(\tau^{'})U(\tau^{'},\tau_{i})d\tau^{'}.
\end{equation}

In the weak coupling regime, the evolution operator can be 
expanded in power series of the interaction Hamiltonian.
To first order, it is given by

\begin{equation}
U(\tau_{f},\tau_{i})=1-i\int_{\tau_{i}}^{\tau_{f}}d\tau^{'}
H_{int}(\tau^{'}).
\end{equation}

The probability amplitude of the transition from the initial 
state $\left|{\cal T}_{i}\right>
=\left|j\right>\otimes\left|\Phi_{i}\right>$ at the 
hypersurface $\tau=0$ to 
$\left|j^{'}\right>\otimes\left|\Phi_{i}\right>$ at $\tau$ is given by

\begin{equation}
\left<j^{'}\Phi_{f}\right|U(\tau,0)\left|j\Phi_{i}\right>
=-i\lambda_{1}\int_{0}^{\tau}d\tau^{'}\left<j^{'}\Phi_{f}\right|d(\tau^{'})
\varphi(x(\tau^{'}))\left|j\Phi_{i}\right>,
\end{equation}

where $\left|\Phi_{f}\right>$ is an arbitrary state of the field and 
$\left|j^{'}\right>$ is the final state of the detector.

The probability of the detector being excited at the 
hypersurface $\tau$, assuming that the detector was prepared 
in the ground state is:

\begin{equation}
P_{eg}(\tau)=\lambda_{1}^{2}|\left<e\right|d(0)\left|g\right>|^{2}
\int_{0}^{\tau}d\tau^{'}\int_{0}^{\tau}d\tau^{''}
e^{iE(\tau^{''}-\tau^{'})}\left<\Phi_{i}\right|\varphi(x(\tau^{'}))
\varphi(x(\tau^{''}))\left|\Phi_{i}\right>,
\end{equation}
where $\omega_{e}-\omega_{g}=E$ is the energy gap between the eigenstates 
of the detector.

Note that we are interested in the final state of the detector
and not that of the field, so we sum over all the possible
final states of the field $\left|\Phi_{f}\right>$. Since the states are 
complete, we have

\begin{equation}
\sum_{f}\left|\Phi_{f}\right>\left<\Phi_{f}\right|=1.
\end{equation}
Eq.(10) shows us that the probability of excitation 
is determined by an integral transform of the positive Wightman function.

Before starting to analyze radiative processes, we would 
like to point out
that a more realistic model of detector must also have 
a continuum of 
states. This asumption allows us to use a first 
order perturbation
theory without taking into account higher order corrections. 
Although
we will use in this paper the two-state model, the case of a 
mixing between a discrete and a continuum 
eigenstates deserves further investigations.

The radiative processes of the uniformly accelerated 
detector discussed 
from the inertial frame point of view was analysed by 
Kolbenstvedt and also 
Grove\cite{Kol}.
Let us analyze the radiative processes of the uniformly 
accelerated Unruh-DeWitt detector from the point of view of 
a observer in a frame comoving with the detector. This approach 
was used by Ginzburg and Frolov \cite{Ginzburg} and 
also Takagi \cite{Takagi}. For the
sake of simplicity we study the two-dimensional case $(D=2)$.

In a $D=2$ dimensional spacetime,
the Rindler coordinates $(\eta,\xi)$ are given by

\begin{equation}
x^{0}=\xi\,sinh\,\eta\qquad\qquad -\infty<\eta<\infty
\end{equation}
\begin{equation}
x^{1}=\xi\,cosh\,\eta\qquad\qquad 0<\xi<\infty.
\end{equation}
where $x^{0}$ and $x^{1}$ are the cartesian coordinates used 
by inertial observers.
The line element in Rindler spacetime is given by

\begin{equation}
ds^{2}=\xi^{2}d\eta^{2}-d\xi^{2}.
\end{equation}

The Rindler edge is globally hyperbolic and 
posesses a timelike Killing
vector which generates a boost about the origin. 
Therefore it is possible to
define positive and negative modes in a unambiguous way. These modes that
form
a complete set, basis 
in the space of the solutions of the massive Klein-Gordon equation 
are given by 

\begin{equation}
\phi_{\nu}(\eta,\xi)=\frac{1}{\pi}(sinh\pi\nu)^{1/2}
e^{-i\nu\eta}K_{i\nu}(m\xi),
\end{equation}

\begin{equation}
\phi_{\nu}^{\ast}(\eta,\xi)=\frac{1}{\pi}(sinh\pi\nu)^{1/2}
e^{i\nu\eta}K_{i\nu}(m\xi),
\end{equation}

where $K_{i\nu}$ is the Bessel function of imaginary 
order or the Macdonald's function \cite{Lebedev}, and $m$ is the 
mass of the field.

In order to canonically quantize the scalar field in the Rindler 
spacetime, let us follow Fulling \cite{Fulling} and
Sciama, Candelas and Deutsch \cite{Sciama}. 
Therefore we expand the field 
operator in the form

\begin{equation}
\varphi(\eta,\xi)=\int^{\infty}_{0}d\nu[b(\nu)\phi_{\nu}
(\eta,\xi)+b^{\dagger}(\nu)\phi^{\ast}_{\nu}(\eta,\xi)],
\end{equation}
where the anihilation operator for Rindler's particle satisfies

\begin{equation}
b(\nu)\left|0,R\right>=0\qquad\qquad    \forall\nu.
\end{equation}

Let us suppose that in the surface $\eta_{i}=constant$, the state of the 
system is $\left|g\right>\otimes\left|0,M\right>$. 
The probability amplitude for the 
system to go to $\left|e\right>\otimes\left|\Phi_{f}\right>$
 at $\eta_{f}$ (where 
$\left|\Phi_{f}\right>$ is any final state of the field) is 

\begin{equation}
A_{|g>\otimes|0,M>\rightarrow|e>\otimes|\Phi_{f}>}=
-i\lambda_{1}d_{eg}(\eta_{i})\int^{\eta_{f}}_{\eta_{i}}
d\eta e^{iE\eta}\left<\Phi_{f}\right|\varphi(\eta,\xi)\left|O,M\right>.
\end{equation}

For a better understanding of the radiative 
processes, let us split the field operator into 
positive and negative parts, i.e.,

\begin{equation}
\varphi(\eta,\xi)=\varphi^{(+)}(\eta,\xi)+\varphi^{(-)}(\eta,\xi),
\end{equation}

where $\varphi^{(+)}(\eta,\xi)$ and $\varphi^{(-)}(\eta,\xi)$ 
are given by:
 
\begin{equation}
\varphi^{(+)}(\eta,\xi)=\int^{\infty}_{0}d\nu b(\nu)
\phi_{\nu}(\eta,\xi)
\end{equation}

and

\begin{equation}
\varphi^{(-)}(\eta,\xi)=\int^{\infty}_{0}d\nu b^{\dagger}(\nu)
\phi^{\ast}_{\nu}(\eta,\xi).
\end{equation}

Substituting eq.(21) and (22) in eq.(19) yields:

\begin{eqnarray}
A_{|g>\otimes|0,M>\rightarrow|e>\otimes|\Phi_{f}>}=
-i\lambda_{1}d_{eg}(\eta_{i})\int^{\eta_{f}}_{\eta_{i}}
d\eta e^{iE\eta}&[&\left<\Phi_{f}
\right|\varphi^{(+)}(\eta,\xi)\left|O,M\right>\nonumber\\ &+&\left<\Phi_{f}\right|\varphi^{(-)}(\eta,\xi)\left|O,M\right>]. 
\end{eqnarray}

The first term is an absorption process and since it contains a factor
of the form $e^{i(E-\nu)\eta}$, it will allow energy to be 
conserved in the asymptotic limit $\eta_{i}\rightarrow-\infty,
\eta_{f}\rightarrow\infty$. The second termis an emission process and 
lead to integrand of the form $e^{i(E+\nu)\eta}$. For $E>0$ and
$\nu>0$ it is rapidly oscilating, giving a negligible contribution.
This expresses the fact that the detector asymptotically can only
suffer a transition to the excited state
absorbing a Rindler particle from the Minkowski 
vacuum $\left|O,M\right>$.
In other words, the detector registers the occupation
number of Rindler particles in the 
Minkowski vacuum. This result allows us
to assume the rotating wave approximation (RWA) to discuss the 
radiative processes\cite{Allen}. This kind of detector is called in the literature 
a square-law detector. If we assume the normally ordered 
field correlation function with respect to the cartesian time, 
the rate at which quanta of 
the field are detected by an inertial detector in the 
Minkowski vacuum state 
vanishes. Since we are interested to study the Unruh-Davies effect from 
the point of view of the accelerated observer we assume the RWA 
with respect to the Rindler time. It is 
important to stress that since the excitation should correspond only 
to the absorption of a Rindler particle the RWA gives exact 
results. As we will see, the RWA allow us to discuss radiative 
processes of an accelerated detector with different proper 
acceleration in both, the begining and the end of the 
observation time.
Using the RWA the probability amplitude becomes:

\begin{equation}
A_{|g>\otimes|0,M>\rightarrow|e>\otimes|\Phi_{f}>}=
-i\lambda_{1}d_{eg}(\eta_{i})\int^{\eta_{f}}_{\eta_{i}}
d\eta e^{iE\eta}\left<\Phi_{f}\right|
\varphi^{(+)}(\eta,\xi)\left|O,M\right>.
\end{equation}

For a better understanding of the RWA and to present 
the difference in the
probability of transition due to it, 
 let us first study the probability of transition without the RWA.
In this case, let us define $F(e,\eta_{i},\eta_{f})$ such that

\begin{equation}
P(E,\eta_{i},\eta_{f},\xi,\xi')=\lambda^{2}_{1}|d_{eg}(\eta_{i})|^{2}
F(e,\eta_{i},\eta_{f},\xi,\xi').
\end{equation}

$F(E,\eta_{i},\eta_{f},\xi,\xi')$ is the response function, i.e., the 
probability of transition normalized by the 
$\lambda^{2}_{1}|m_{eg}(\eta_{i})|^{2}$ term. 
Thus 

\begin{equation}
F(E,\eta_{i},\eta_{f},\xi,\xi')=
\int^{\eta_{f}}_{\eta_{i}}d\eta\int^{\eta_{f}}_{\eta_{i}}d\eta^{'}
e^{iE(\eta-\eta^{'})}
\left<0,M\right|\varphi
(\eta^{'},\xi^{'})\varphi(\eta,\xi)\left|0,M\right>.
\end{equation}

Note that  
in the calculations we are not restricted to constant proper 
acceleration of the detector but we are taking into account the 
general case of non-constant proper acceleration of the detector
during the measurement process. 

Following Svaiter and Svaiter \cite{Nami1}, let us define 

\begin{equation}
\eta-\eta{'}=\tau
\end{equation}

and

\begin{equation}
\eta_{f}-\eta_{i}=\Delta T.
\end{equation}

We would like to stress that Levin, Peleg and Peres 
\cite{Peres} also used the same 
technique to study radiative processes in finite observation times.
Substituting eqs.(27) and (28) in eq.(26) we have

\begin{equation}
F(E,\Delta T, \xi,\xi')=
\int^{\Delta T}_{-\Delta T}d\tau(\Delta T-|\tau|)
e^{iE\tau}
\left<0,M\right|\varphi(\eta^{'},\xi^{'})
\varphi(\eta,\xi)\left|0,M\right>.
\end{equation}

Let us define the rate $R(E,\Delta T,\xi,\xi')$, i.e., 
the probability transition per unit time, as :
 
\begin{equation}
R(E, \Delta T,\xi,\xi')=\frac{dF(E,\Delta T,\xi,\xi')}{d(\Delta T)}
\end{equation}

Consequently we have:

\begin{equation}
R(E,\Delta T,\xi,\xi')=
\int^{\Delta T}_{-\Delta T}d\tau
e^{iE\tau}
\left<0,M\right|\varphi(\eta^{'},\xi^{'})\varphi
(\eta,\xi)\left|0,M\right>.
\end{equation}

This important result shows that asymptotically the rate of 
excitation of the detector is given by the Fourier transform 
of the positive frequency 
Wightman function. This is exactly the quantum version of the 
Wiener-Khintchine theorem which asserts that the spectral density 
of a stationary random variable is the Fourier transform of the 
two point-correlation function.

The rate rigorously is:

\begin{eqnarray}
R(E,\Delta T,\xi,\xi')=
\int^{\Delta T}_{-\Delta T}d\tau
e^{iE\tau}&[&
\left<0,M\right|\varphi^{(+)}(\eta^{'},\xi^{'})
\varphi^{(+)}(\eta,\xi)\left|0,M\right>\nonumber\\ &+&
\left<0,M\right|\varphi^{(-)}(\eta^{'},\xi^{'})
\varphi^{(-)}(\eta,\xi)\left|0,M\right> + 
\left<0,M\right|\varphi^{(-)}(\eta^{'},\xi^{'})
\varphi^{(+)}(\eta,\xi)\left|0,M\right>\nonumber\\ &+&
\left<0,M\right|\varphi^{(+)}(\eta^{'},\xi^{'})
\varphi^{(-)}(\eta,\xi)\left|0,M\right>].
\end{eqnarray}

The last matrix element can be writen as                 

\begin{eqnarray}
\left<0,M\right|\varphi^{(+)}(\eta',\xi')
\varphi^{(-)}(\eta,\xi)\left|0,M\right>&=&
\left<0,M\right|\varphi^{(-)}(\eta,\xi)
\varphi^{(+)}(\eta',\xi')\left|0,M\right>\nonumber\\ &+&
[\varphi^{(+)}(\eta',\xi'),\varphi^{(-)}(\eta,\xi)] .
\end{eqnarray}

The commutator is a c-number independent of the initial 
state of the field. Many authors in quantum optics 
claim that this contribution has no great physical 
interest. So the matrix elements determining the 
detection of quanta of the field are of the form 

\begin{equation}
\left<0,M\right|\varphi^{(-)}(\eta^{'},\xi^{'})
\varphi^{(+)}(\eta,\xi)\left|0,M\right>
 + \left<0,M\right|\varphi^{(-)}(\eta,\xi)
\varphi^{(+)}(\eta',\xi')\left|0,M\right>.
\end{equation}

Such approximation is called the rotating-wave approximation (RWA).
The last expression said that the excitation of the detector 
corresponds to the absorption of a Rindler particle from the 
Minkowski vacuum $\left|0,M\right>$. This is a consequence of the 
absorptive nature of the detector. 
Nevertheless there are subtleties in the process.
As it was shown by Unruh and Wald \cite{Unruh2}, the 
process is followed 
by the emission of a Rindler particle in a causally 
disconected region 
of the spacetime. (In the appendix we discuss this result).
This result can be understood since the Minkowski vacuum
 $\left|0,M\right>$
can be expressed into a set of EPR type of Rindler particles
 \cite{Pringle}.
Thus a coherent state with respect to the annihilation operator
 of Minkowski 
particles appears to be squeezed with respect to the annihilation 
operator of Rindler particles. Note that this fact has been 
investigated in the literature
in the context of quantum optics. We can imagine 
macroscopic detectors 
different from the square law detector. For example, we can 
use the process of stimulated emission as a basis for detection.
In this case the field operator would occur in anti-normal order.
Different operator ordering has been also sugested. 
Wilkens and Lewenstein 
\cite{Wilkens}
proposed a photodetection scheme based on an interference 
between emission and absorption processes. In this 
paper we are assuming the absorptive nature of the detector.

Another point which is important to stress 
is the fact that a logarithmic ultraviolet
divergence will appear in the response function, as was
discussed by Svaiter and Svaiter \cite{Nami1}. In order to
circumvent the problem of the divergence, Higuchi et al
\cite{Higuchi} considered a detector switched on and off
continuously with the field. These authors claim that the
ultraviolet logarithmic divergence in the excitation
probability that Svaiter and 
Svaiter \cite{Nami1} found in the time
dependent perturbation theory can be circumvented.
Defining the probability of transition 
per unit proper time as 
$\frac{dF(E,\Delta T)}{d\Delta T}$, normalized 
by the selectivity of the detector, it is possible to
obtain the 
rate of spontaneous excitation after a finite 
observation or switching time. 
Although the rate of 
spontaneous excitation of an 
unaccelerated atom (assuming that the 
state of the field is the Minkowski vacuum state 
and the state of the 
detector is the ground state) is negative and diverges
 below the uncertainty region, i.e. 
for $\Delta T\leq \frac{1}{E}$, there is no problem in 
such behavior. This happens because it is possible to consider 
measurements of finite 
duration only for $\Delta T > \frac{1}{E}$. Over shorter 
time intervals, 
it is not even
 possible to say what level our two-level system is in, or even 
to define this two-level system. Note that we are 
not interested in 
discussing the subtle problem of
 how to decode the information stored in the system and to 
convert it into a classical signal. Only with the latter 
the measurement process 
is complete. Without this mechanism it is 
convenient to call the first step 
as a "pre-measurement", but because we are not interested 
in discussing this 
controversial issue, we will still call the first step as a 
measurement. Back to our problem, we conclude 
that measurement 
is meaningless for intervals obeying $ \Delta T\leq \frac{1}{E}$. 
As it was shown by Ford, Svaiter and Lyra \cite{Nami2}, 
the asymptotic regime is achieved after a very 
short transient 
period. We conclude that there is no problem 
with the rate, 
concerning non positive definite rate, or 
divergent in intervals above the uncertainty region.

Of course, this divergence is expected, in the 
sense that the response function is an integral transform of 
the positive 
Wightman function, which becomes singular at one point of 
spacetime. Since 
the field is a distribution, the square of
 such object in one point of spacetime 
is ill-defined. This is
the fundamental problem of the interacting field theory in 
flat and curved 
spacetime. If we adopt the point of view that we are 
interested in 
$measurements$, everything is in order and therefore 
renormalization procedure is not required. 
Sriramkumar and Padmanabhan obtained a different conclusion 
\cite{Sri} however, with we disagree.
Note that we can 
use another 
physical interpretation to the mathematical calculations, as the 
result of making two measurements in the system separated 
by the time interval $\Delta\tau$, instead of the usual 
switching interpretation.  For a short time interval, 
we would 
expect a large 
disturbance in the system. It is clear that the 
effect of a smooth
 switching on and off, sugested by Higuchi and 
collaborators (as the more realistic 
procedure for modeling detectors) 
is to prevent the region $ 
\Delta\tau\leq \frac{1}{E}$. Again 
we are not interested in discussing 
another controversial issue: the uncertainty 
relation for time and energy and the different interpretations 
of such 
inequality and we are adopting the Landau and Peierls 
approach \cite{Landau}.  
In this case the energy uncertainty introduced by the 
switching is
less than the level separation of the system and the 
measurement can
be defined.

 The conclusion is that the transient terms related with the 
switching 
or finite observational time will 
vanish in the asymptotic limit. A straightforward
calculation gives the transition 
probability per unit proper time (normalized by 
the selectivity 
of the atom) 
of an inertial atom interacting 
with a {\it massless} 
field in the Minkowski vacuum (without  assuming the RWA) in a four 
dimensional spacetime:  

\begin{equation}
R(E,\Delta T)=\frac{1}{2\pi}
\biggl(-E\Theta(-E)+\frac{ cosE\Delta T}{\pi
\Delta T}+\frac{|E|}{\pi}(Si|E|\Delta T-\frac{\pi}{2})\biggr),
\end{equation}
where $Si(z)$ is the 
sine integral function defined by 
$Si(z)=\int^{\infty}_{0}\frac{sint}{t}dt$
\cite{Lebedev}. For large values of the argument we 
have $Si(\infty)=\frac{\pi}{2}$, thus
in the asymptotic limit, we will obtain for the rate  

\begin{equation}
lim_{\Delta T\rightarrow\infty}R(E,\Delta T)=
\frac{-E}{2\pi}\Theta(-E).
\end{equation} 

The transition rate is proportional to the energy level gap 
of the atom, and in the asymptotic limit only spontaneous decay is 
allowed.

It is possible to repeat the calculations for the uniformly accelerated 
detector. In the 
asymptotic limit the 
 expression that takes into account 
the two processes: spontaneous and induced decay or induced 
excitation is 
given by:

\begin{equation}
lim_{\Delta\tau\rightarrow\infty}
R(E,\Delta\tau,\alpha)=\frac{|E|}{2\pi}\biggl(\Theta(-E)
(1+\frac{1}{e^{2\pi\alpha |E|}-1})+\Theta(E)
\frac{1}{e^{2\pi\alpha E}-1}\biggr).
\end{equation}
were $\frac{1}{\alpha}$ is the proper acceleration of the atom.

Let us use the RWA to obtain the probability of transition per unit 
time in the general case of a {\it massive} field in $D=2$.
Note that no infrared divergences should appear since the 
field is massive. For the ultraviolet divergences the response function
per unit time is finite for $\Delta T> \frac{1}{|E|}$

Going back to the eq.(32) using the RWA we have to the rate:

\begin{eqnarray}
R(E,\Delta T,\xi,\xi')=
\int^{\Delta T}_{-\Delta T}d\tau
e^{iE\tau}&[&
\left<0,M\right|\varphi^{(-)}(\eta^{'},\xi^{'})
\varphi^{(+)}(\eta,\xi)\left|0,M\right>\nonumber\\ 
&+&
\left<0,M\right|\varphi^{(-)}(\eta,\xi)
\varphi^{(+)}(\eta',\xi')\left|0,M\right>].
\end{eqnarray}

Defining $v(\nu,\xi)$ and the complex 
conjugate  $v^{\ast}(\nu,\xi)$
such that

\begin{equation}
\phi_{\nu}(\eta,\xi)=
e^{-i\nu\eta}v(\nu,\xi)
\end{equation}
and
\begin{equation}
\phi_{\nu}^{\ast}(\eta,\xi)=
e^{i\nu\eta}v^{\ast}(\nu,\xi),
\end{equation}

we have 

\begin{eqnarray}
R(E,\Delta T,\xi,\xi')&=&
\int^{\Delta T}_{-\Delta T}d\tau
e^{iE\tau}\int d\sigma d\nu 
\left<0,M\right|b^{\dagger}(\sigma)b(\nu)
\left|O,M\right>\nonumber\\
&[&e^{i(\sigma\eta'-\nu\eta)}
v^{\ast}(\sigma,\xi')v(\nu,\xi)+e^{i(\sigma\eta-\nu\eta')}
v^{\ast}(\sigma,\xi)v(\nu,\xi')].
\end{eqnarray}

Using the fact that

\begin{equation}
\left<0,M\right|b^{\dagger}(\sigma)b(\nu)\left|O,M\right>= \frac{\delta(\sigma-\nu)}
{e^{2\pi\sigma}-1},
\end{equation}

we have the rate

\begin{eqnarray}
R(E,\Delta T,\xi',\xi)=\int^{\infty}_{0}d\nu&[&
\frac{1}{e^{2\pi\nu}-1}v^{\ast}(\nu,\xi')v(\nu,\xi)
\int^{\Delta T}_{-\Delta T}d\tau e^{i(E-\nu)\tau}\nonumber\\
&+&\frac{1}{e^{2\pi\nu}-1}v^{\ast}(\nu,\xi)v(\nu,\xi')
\int^{\Delta T}_{-\Delta T}d\tau e^{i(E+\nu)\tau}]. 
\end{eqnarray}

Note the absence of the first term of eq.(37) in eq.(43). 
The spontaneous emission term does not appear since we 
adopted the RWA and disregard the commutator 
$[\varphi^{+}(\eta',\xi'),\varphi^{-}(\eta,\xi)]$
in the rate $R(E,\Delta T)$. The conclusion is that 
this term (which is independent of the state of the 
field) is responsible for the spontaneous emission processes
(this radiative process is induced by vacuum fluctuations).
It is well known that in a semiclassical theory where a 
quantum mechanical system interacts with a normal ordered field only 
stimulated emission and absortion are predicted. The use of the 
RWA or the Glauber\cite{Glauber} correlation function possesses 
a simple physical interpretation of a field with 
classical correlation functions interacting with the detector.

As it has been pointed out by Milonni and 
Smith \cite{Milonni} and Ackerhalt,
Knight and Eberly \cite{Knight}, there is a different 
approach to study
radiative processes without using perturbation theory, 
but using the 
Heisenberg equations of motion. In this approach it is 
possible to
obtain nonperturbative expressions for radiative processes 
where the
radiation reaction appears in a very simple way: 
the part of the field 
due to the atom (detector) that drives the Dicke 
operators \cite{Dicke}.
In this approach it is possible to identify the role 
of radiation reaction
and vacuum fluctuations in spontaneous emission. We 
would like to stress 
the fact that the contribution of vacuum fluctuations 
and radiation 
reaction can be chosen arbitrarily, depending on the 
order of the Dicke 
and field operators. As it was discussed by Dalibard, 
Dupont-Roc and 
Cohen-Tannoudji \cite{Cohen}, there is a preferred ordering 
in such a
way that the vacuum fluctations and radiation reaction 
contribute
equally to the spontaneous emission process. More 
recently this approach
was developed by Audretsch and Muller, Audretsch, Mensky and 
Muller and also 
Audretsch, Muller
and Holzmann \cite{Audretsch 1} to study the 
Unruh-Davies 
effect.
These authors constructed the following picture of the 
Unruh-Davies effect.
The effect of vacuum fluctuations is changed by the 
acceleration, 
although the contribution of radiation reaction is unaltered.
Due to the modified vacuum fluctuation contribution, 
transition to an
excited state becomes possible even in the vacuum.
Note that different results appear in the literature. 
Barut and Dowling 
\cite{Barut}  concluded that the thermal response of 
the detector does
not arise through
an interaction with "real" particles, but from the 
spectrum of its 
self-field which has become altered by the change to a 
non-inertial
frame. The discrepancy between above cited results
comes from the arbitrariness in the operator ordering when 
one constructs
the interaction Hamiltonian.

It is important to compare our results with the conclusions
obtained by Padmanabhan and Singh \cite{Pad}. In the above mentioned
article, these authors presented an example (a detector in 
a uniformly
rotating frame) to conclude that the monopole detector is a 
fluctuometer 
\cite{Candelas}. In this case, although the expectation value of the 
number operator (in the rotating frame) in the Minkowski 
vacuum is zero,
the power spectrum (the Fourier transform of the positive 
Wightman function)
is not zero. They concluded that the power spectrum is not 
related to the existence of "real" particles. The uniformly 
accelerated detector 
goes to the excited state not because there is real Rindler particles
in the Mikowski vacuum, but because the accelerating source supplies
energy to the transition.

The key point of this discussion is given by Eq.(31) and Eq.(33). 
Assuming the RWA we disregard a vacuum piece, i.e., a contribution 
that is independent of the state of the field. Nevertheless 
we still have 
obtained a non-zero probability 
of excitation, since the Minkowski vacuum is filled by 
thermal Rindler particles. Eq.(38) gives to the 
power spectrum
associated: absorption of
a Rindler particle. In other words, assuming the RWA the power 
spectrum and the expectation value of the Rindler number operator
in the Minkowski vacuum must be proportional. The inclusion of the 
commutator of Eq.(33) in Eq.(38) gives a vacuum piece 
contribution. 

Actually, the puzzle of the rotating detector \cite{Letaw}
was solved a few months ago by Davies, Dray and Manogue \cite{Davies}. 
These authors assumed that the field is defined only in the 
interior of a cilinder of radius $a$ in such a way that the 
rotating Killing vector $\partial_{t}-\Omega\partial_{\theta}$ is 
always timelike. Consequently the response function of the rotating 
detector is zero. Of course if the angular velocity of the 
detector is above some threshold, excitation occurs. Clearly the 
excitation of the rotating detector is related with the  "Galilean" transformation to 
rotating coordinates which is not defined above certain radius. 
It is interesting to study the following situation: take a coordinate 
transformation between the inertial and the rotating coordinates 
defined by Takeno \cite{Takeno}. The advantage of this coordinate 
transformation is that the velocity of a rotating point is $v=\tanh \Omega r$
(for small radius or angular velocities we recovered the situation 
$v=\Omega r$). This coordinate transformation cover all the Minkowski 
manifold for all angular velocities. We conjecture that in this 
case the answer obtained calculating the Bogoliubov coefficients 
between cartesian and rotating modes and the response function of the 
detector will agree.

Going back to eq.(43), for large values of $\Delta T$ we have

\begin{equation}
lim_{\Delta T\rightarrow\infty}R(E,\Delta T,\xi,\xi')=
\frac{1}{\pi}cosh\frac{\pi E}{\Delta T}\int_{-\infty}^{\infty}
d\lambda e^{iE \lambda} K_{0}(m\sigma),
\end{equation}

where $K_{0}$ is the Bessel Neumann function, and

\begin{equation}
\sigma^{2}=(\xi^{2}+\xi'^{2}+2\xi\xi' cosh\lambda).
\end{equation}

Finally note that we did all the calculations using the Rindler's
time $\eta$, and the proper acceleration does not appear in Eq.(46).
To obtain the correct expression we have to relate the Rindler
time to the detector's proper time. Using this fact we 
have to do the
replacement

$$
\nu\rightarrow 2\pi\alpha\nu
$$
where $\alpha^{-1}$ is the proper acceleration of the 
detector. With this 
replacement we recovered the usual result.

\section{Stress Tensor Detector}

In this section we will repeat all the calculations that 
we did in the
previous section, using the following interaction 
Lagrangean density:

\begin{equation}
L_{int}=\lambda_{2}\sum_{j=1}^{n}
d_{j}^{\mu}(\tau)\partial_{\mu}\varphi(x).
\end{equation}

where $d_{j}^{\mu}=m(\tau) l_{j}^{\mu}$, and $l_{j}^{\mu}$
is a set of vectors.

With this construction it is possible to construct a 
non-gravitational stress-tensor detector, by combining the
response function of different detectors, as was discussed
by Ford and Roman \cite{Ford}. For large $\Delta T$ this 
detector is still a particle detector with a derivativelly 
coupled Lagrangian. For short time intervals $(\Delta T\rightarrow 0)$ 
the detector will measure the stress-tensor of the field. In this 
last situation, as was discussed these various detectors will
measure a particular component of the stress-tensor of a 
massive minimal coupled scalar field.

The calculations are exactly the same as we did. The only difference is
that we have the components $R_{00}$, $R_{01}$ and $R_{11}$. The rate
in a uniformly accelerated world-line is given by 

\begin{equation}
R(E,\Delta T,\xi,\xi)=\lambda_{2}^{2}
\sum_{i,j=1}^{n}<d_{j}^{\mu}><d_{i}^{\nu}>
R_{\mu\nu}(E,\Delta T,\xi),
\end{equation}

where 

\begin{eqnarray}
R_{00}(E,\Delta T,\xi)=\int^{\infty}_{0}d\nu&[&
\frac{\nu^{2}}{e^{2\pi\nu}-1}v^{\ast}(\nu,\xi')v(\nu,\xi)
\int^{\Delta T}_{-\Delta T}d\tau e^{i(E-\nu)\tau}\nonumber\\
&+&\frac{\nu^{2}}{e^{2\pi\nu}-1}v^{\ast}(\nu,\xi)v(\nu,\xi')
\int^{\Delta T}_{-\Delta T}d\tau e^{i(E+\nu)\tau}] ,
\end{eqnarray}

\begin{eqnarray}
R_{01}(E,\Delta T,\xi)=\int^{\infty}_{0}d\nu&[&
\frac{\nu}{e^{2\pi\nu}-1}
v^{\ast}(\nu,\xi')\frac{\partial}{\partial\xi}v(\nu,\xi)
\int^{\Delta T}_{-\Delta T}d\tau e^{i(E-\nu)\tau}\nonumber\\
&+&\frac{\nu}{e^{2\pi\nu}-1}\frac{\partial}{\partial\xi}
v^{\ast}(\nu,\xi)v(\nu,\xi')
\int^{\Delta T}_{-\Delta T}d\tau e^{i(E+\nu)\tau}] ,
\end{eqnarray}

and

\begin{eqnarray}
R_{11}(E,\Delta T,\xi)=\int^{\infty}_{0}d\nu&[&
\frac{1}{e^{2\pi\nu}-1}\frac{\partial}{\partial\xi'}v^{\ast}(\nu,\xi')
\frac{\partial}{\partial\xi}v(\nu,\xi)\int^{\Delta T}_{-\Delta T}
d\tau e^{i(E-\nu)\tau}\nonumber\\
&+&\frac{1}{e^{2\pi\nu}-1}\frac{\partial}
{\partial\xi}v^{\ast}(\nu,\xi)
\frac{\partial}{\partial\xi'}v(\nu,\xi')
\int^{\Delta T}_{-\Delta T}d\tau e^{i(E+\nu)\tau}]. 
\end{eqnarray}

Substituting eq.(15) and (16) in eq.(48), (49) and (50),
we have after a straightforward calculation

\begin{equation}
R_{00}(E,\Delta T,\xi)=\frac{\Delta T}{\pi^{2}}
\int_{0}^{\infty}d\nu \nu^{2} f(\nu,E,\Delta T)
K_{i\nu}(m\xi)K_{i\nu}(m\xi),
\end{equation}

\begin{equation}
R_{01}(E,\Delta T,\xi)=\frac{m\Delta T}{\pi^{2}}
\int_{0}^{\infty}d\nu \nu f(\nu,E,\Delta T)
K_{i\nu}(m\xi)\frac{\partial}{\partial\xi}K_{i\nu}(m\xi),
\end{equation}

\begin{equation}
R_{11}(E,\Delta T,\xi)=\frac{m^{2}\Delta T}{\pi^{2}}
\int_{0}^{\infty}d\nu  f(\nu,E,\Delta T)\frac{\partial}{\partial\xi}
K_{i\nu}(m\xi)\frac{\partial}{\partial\xi} K_{i\nu}(m\xi),
\end{equation}

where

\begin{equation}
f(E,\Delta T,\nu)=e^{-\pi\nu}
\left(\frac{(sin(E-\nu)\Delta T}{(E-\nu)\Delta T} +
\frac{(sin(E+\nu)\Delta T}{(E+\nu)\Delta T}\right).
\end{equation}

For small proper acceleration or $m\xi>>1$, we can use the
asymptotic expansion for large arguments of the Bessel function 
$K_{i\nu}(m\xi)$

\begin{equation}
K_{i\nu}(m\xi)=(\frac{\pi}{2m\xi})^{1/2} e^{-m\xi}
(1-\frac{(1+4\nu^{2})}{8m\xi}),
\end{equation}

and

\begin{equation}
\frac{\partial}{\partial\xi}K_{i\nu}(m\xi)=
-(\frac{\pi}{2m\xi})^{1/2} e^{-m\xi}
(1+\frac{(3-4\nu^{2})}{8m\xi}).
\end{equation}

Substituting eq.(55) and (56) in eqs.(51), (52) and (53) we have

\begin{equation}
R_{00}(E,\Delta T,\xi)=\frac{\Delta T}{\pi^{2}}
\int_{0}^{\infty}d\nu  f(\nu,E,\Delta T)
(\alpha_{1}(m\xi)\nu^{2}+\alpha_{2}(m\xi)\nu^{4}+
\alpha_{3}(m\xi)\nu^{6}),
\end{equation}

\begin{equation}
R_{01}(E,\Delta T,\xi)=-\frac{m\Delta T}{\pi^{2}}
\int_{0}^{\infty}d\nu  f(\nu,E,\Delta T)
(\beta_{1}(m\xi)\nu+\beta_{2}(m\xi)\nu^{3}+
\beta_{3}(m\xi)\nu^{5}),
\end{equation}

\begin{equation}
R_{11}(E,\Delta T,\xi)=\frac{m^{2}\Delta T}{\pi^{2}}
\int_{0}^{\infty}d\nu  f(\nu,E,\Delta T)
(\gamma_{1}(m\xi)+\gamma_{2}(m\xi)\nu^{2}+
\gamma_{3}(m\xi)\nu^{4}),
\end{equation}

where

\begin{equation}
\alpha_{1}=1-\frac{1}{4m\xi}+\frac{1}{64m^{2}\xi^{2}}
\end{equation}

\begin{equation}
\alpha_{2}=-\frac{1}{m\xi}+\frac{1}{8m^{2}\xi^{2}}
\end{equation}

\begin{equation}
\alpha_{3}=\frac{1}{m^{2}\xi^{2}}
\end{equation}

\begin{equation}
\beta_{1}=1-\frac{1}{4m\xi}-\frac{3}{64m^{2}\xi^{2}}
\end{equation}

\begin{equation}
\beta_{2}=-\frac{1}{m\xi}-\frac{1}{8m^{2}\xi^{2}}
\end{equation}

\begin{equation}
\beta_{3}=\frac{1}{4m^{2}\xi^{2}}
\end{equation}

\begin{equation}
\gamma_{1}=1-\frac{3}{4m\xi}+\frac{9}{64m^{2}\xi^{2}}
\end{equation}

\begin{equation}
\gamma_{2}=-\frac{1}{m\xi}+\frac{3}{8m^{2}\xi^{2}}
\end{equation}

\begin{equation}
\gamma_{3}=\frac{1}{4m^{2}\xi^{2}}
\end{equation}

Note the presence of a smooth cutoff $e^{-\pi\nu}$ that prevents 
the existence of ultraviolet divergences. Another important point 
is that it is possible negative spontaneous rate of transition. A
negative value simply means that the detector's rate is less than 
the rate with the last term of eq.(33) (the commutator) included. 
This is the vacuum contribution part. Consequently using the 
RWA we are subtracting the detector's response in the vacuum state 
as did Ford and Roman \cite{Ford}. In the next section we will discuss the 
response function of a
detector which is inertial in the infinite past and has a
constant proper acceleration in the infinite future.

\section{The asymptotic accelerated detector}

The aim of this section is to discuss the following physical 
situation. How the monopole detector behaves if is traveling 
along a world line in such a way the detector is inertial in the 
infinite past and has a constant proper acceleration in the 
infinite future.

Let us consider the folowing transformation of coordinates 
between the inertial $(t,x)$ and non-inertial 
coordinates $(\eta,\xi)$,

\begin{equation}
t+x=\frac{2}{a}sinh a(\eta+\xi)
\end{equation}

\begin{equation}
t-x=-\frac{1}{a}e^{a(\eta+\xi)}
\end{equation}

This coordinate system was investigated by Kalnins 
and Miller\cite{Miller},
and by this reason we will call it as the Kalnins and 
Miller coordinate
system.

The line element in this coordinate system can be 
written as:

\begin{equation}
ds^{2}=(e^{2a\xi}+e^{-2a\eta})(d\eta^{2}-d\xi^{2})
\end{equation}

The proper acceleration in the world line $\xi=cte$ is
given by 

\begin{equation}
lim_{\eta\rightarrow -\infty} \alpha(\eta,\xi_{0})=0
\end{equation}

and

\begin{equation}
lim_{\eta\rightarrow +\infty} \alpha(\eta,\xi_{0})=a e^{-a\xi_{0}}
=\alpha_{\infty}
\end{equation}

The discussion of these issues was given by
ref.\cite{Costa}. Eqs.(72)
and (73) show
that $\xi=cte$ is the world line of an uniformly accelerated observer. 
Note that the hypersurface $\eta=\eta_{0}$ is a Cauchy surface 
for the region $t-x < 0$.

The massive Klein-Gordon equation in the Kalnins-Miller manifold 
reads

\begin{equation}
[\frac{\partial^{2}}{\partial\eta^{2}}- 
\frac{\partial^{2}}{\partial\xi^{2}}+m^{2}(e^{-2a\eta}+e^{2a\xi})]
\varphi(\eta,\xi)=0.
\end{equation}

Since the coordinate system $(\eta,\xi)$ allows the separation 
of variables,
writing $\varphi(\eta,\xi)=F(\eta)G(\xi)$, the Klein-Gordon
equation separates in the following equations:

\begin{equation}
(\frac{d^{2}}{d\eta^{2}}+m^{2}e^{-2a\eta}+a\nu)F(\eta)=0
\end{equation}

and

\begin{equation}
(\frac{d^{2}}{d\xi^{2}}-m^{2}e^{2a\xi}+a\nu)G(\xi)=0.
\end{equation}

Equations(75) and (76) are Bessel equations with imaginary order. Although
we separate the solution in a time dependent part
multiplied by the spatial dependent part, 
the situation is very different from the Rindler coordinate 
system. The metric is not static and there is an ambiguity 
in defining positive and negative frequencies modes. A 
straightforward calculation reveals that there are two 
well behaved complete set $(\phi_{\nu}(x), \phi^{\ast}_{\nu}(x))$ 
and 
$(\varphi_{\nu}(x), \varphi^{\ast}_{\nu}(x))$, basis in the 
space of the solutions of the massive Klein-Gordon equation in the 
Kalnins-Miller manifold. These two complete sets are given by

\begin{equation}
\phi_{\nu}(\eta,\xi)=
\frac{1}{2}\left(\frac{\nu(1-e^{-2\pi\nu})}{\pi a}\right)^{\frac{1}{2}}
H^{(1)}_{i\nu}(\frac{m}{a}e^{-a\eta})K_{i\nu}(\frac{m}{a}e^{a\xi})
\end{equation}

\begin{equation}
\phi^{\ast}_{\nu}(\eta,\xi)=
\frac{1}{2}\left(\frac{\nu(1-e^{-2\pi\nu})}{\pi a}\right)^{\frac{1}{2}}
H^{(2)}_{i\nu}(\frac{m}{a}e^{-a\eta})K_{i\nu}(\frac{m}{a}e^{a\xi})
\end{equation}

and

\begin{equation}
\varphi_{\nu}(\eta,\xi)=(\frac{\nu}{\pi a})^{\frac{1}{2}}
J_{i\nu}(\frac{m}{a}e^{-a\eta})K_{i\nu}(\frac{m}{a}e^{a\xi})
\end{equation}

\begin{equation}
\varphi^{\ast}_{\nu}(\eta,\xi)=(\frac{\nu}{\pi a})^{\frac{1}{2}}
J_{-i\nu}(\frac{m}{a}e^{-a\eta})K_{i\nu}(\frac{m}{a}e^{a\xi}).
\end{equation}

Since the line element is time dependent, there is no simple way 
to define positive and negative frequency modes. Different solutions
for the problem were presented by di Sessa \cite{di} and
Sommerfield \cite{Som}. di Sessa claims that the concept of
positive frequency requires for its definition a complexification
of the Minkowski manifold. In this situation the positive
frequency modes are those which vanish when $t\rightarrow-\infty$.
Then the modes given by eqs.(77) and (78) are positive and negative
frequency
modes in the infinite past. Sommerfield used another criteria to
define positive and negative frequency modes. The dilatation
operator generates translation in time, since it satisfies
the Heisenberg equation of motion 

$$
[\varphi,D]=i\frac{\partial\varphi}{\partial\tau}.
$$

Using this fact and the asymptotic expression for the Bessel
function of the first kind, Sommerfield chose the modes given 
by eqs.(79) and (80)
as positive and negative frequency modes in the infinite future.
By this reason we will call them "inertial" and "accelerated" modes 
respectively.

 The Bogoliubov coefficients between the Minkowski modes
and the inertial and accelerated modes are given respectively by :

\begin{equation}
\left|\beta_{\nu\mu}\right|^{2}_{in}=\frac{1}{2\pi^{2}\nu\epsilon 
sinh\pi\nu}(Re)^{2}{(-i\nu)![2a(\epsilon-\mu)]^{i\nu}}
\end{equation}

and

\begin{equation}
\left|\beta_{\nu\mu}\right|^{2}_{ac}=\frac{1}{2\pi\epsilon a}
\frac{1}{e^{2\pi\nu}-1}
\end{equation}

where $\epsilon$ is the energy of the Minkowski modes. 
Using eq.(81) and 
(82) it is possible to obtain the transition rate of the detector.
The problem is that it is not possible to make asymptotic
measurements, 
but only for finite time $\Delta T=\eta_{f}-\eta_{i}$. For small 
$\Delta T$ it is possible to assume that the metric is static and we 
have two different outcomes if the measurement is made in the infinite
past or in the infinite future. 
The calculation of the response function in the infinite past is
very difficult to evaluate exactly. Nevertheless the response function 
in the infinite future can be evaluated. We will show that the rate
of transition in the infinite future is the same as the Rindler's case.
To obtain this result we have to substitute eq.(79) and (80) in eq.(38) and
use the expression of the Bogoliubov coefficient between the
Minkowski modes and the accelerated modes. Thus using eq.(38) we have

\begin{eqnarray}
R(E,\eta,\eta',\xi,\xi')&=&\int_{-\Delta T}^{\Delta T}
d\tau e^{iE\tau} \int d\eta \frac{1}{e^{2\pi\mu}}\frac{\mu}{(\pi a)^{2}}
(J_{-i\mu}(\frac{m}{a}e^{-a\eta})
J_{i\mu}(\frac{m}{a}e^{-a\eta'})\nonumber\\ &+&
J_{-i\mu}(\frac{m}{a}e^{-a\eta'})
J_{i\mu}(\frac{m}{a}e^{-a\eta}))
K_{i\mu}(\frac{m}{a}e^{a\xi})
K_{i\mu}(\frac{m}{a}e^{a\xi'}).
\end{eqnarray}

Substituting the asymptotic expression for the Bessel function 
of first kind
given by

$$
lim_{x\rightarrow 0} J_{\nu}(x)=\frac{x^{\nu}}{2^{\nu}\Gamma(1+\nu)}
$$

in eq.(83) we have

\begin{equation}
R(E,\Delta T,\xi,\xi')=\frac{\Delta T}{\pi^{2}}
\int_{0}^{\infty}d\mu  f(\mu,E,\Delta T)
K_{i\mu}(m\xi)K_{i\mu}(m\xi'),
\end{equation}

We obtain that if a
measurement is made in the asymptotic future we have a similar result
as Rindler's. Nevertheless if the measurement is made in 
the infinite 
past we obtain a non-expected result, i.e, the rate is 
not zero although
the detector is inertial in this region. This result can be understood 
since although the detector is travelling in an inertial world line 
the coordinate system is non-cartesian.

\section{Conclusions}

In this paper we discussed radiative processes of different detectors
interacting with a massive scalar field. The probability of
transition per unit proper time of the accelerated detector 
is obtained
for the monopole Unruh-DeWitt detector and also for the 
derivatively
coupled detector.                

We used the RWA in order to simplify the calculations of a 
detector
 with nonconstant proper acceleration. We want to remark that 
the standard 
"photodetection" scheme is based on absorption of quantum of
 the field by the detector. Using first order perturbation 
theory the rate of excitation is proportional to the Fourier
 transform of a normal ordered product of the
negative and positive parts of the field operator. Nevertheless, 
normal ordering with respect to the Rindler time is not normal 
ordered with respect
 to the Minkowski time. Therefore an absorption of a Rindler 
particle by the detector is view by an inertial observer by 
an emission and absorption of Minkowski particles, i.e., the 
annihilation and creation operators of
Rindler particles contains a mixture of both positive and negative
 frequency
parts of the field operator with respect to the inertial time.

Summarizing, the Minkowski vacuum  $\left|0,M\right>$ with respect
to the annihilation operator of inertial particles is a squeezed 
state with respect to the annihilation operator of Rindler particles.
In this way, the absorption of Rindler particles combine 
both absorption
and emission of Minkowski particles in the excitation processes.

\section{Appendix}

In the appendix we will demonstrate that the excitation of the 
detector travelling in the right edge of the Rindler manifold is 
followed by a annihilation of a Rindler particle in the right 
edge of the Rindler manifold and the creation of a Rindler 
particle in its left edge.

For the sake of simplicity let us suppose that 
the mass of the quanta of the field is zero i.e. $m^{2}=0$ and define

\begin{equation}
\mbox{}^{R}u_{k}=(4\pi\nu)^{-1/2}e^{i(k\xi-\nu\eta)}\qquad\qquad  in R,
\end{equation}

\begin{equation}
\mbox{}^{R}u_{k}=0 \qquad\qquad   in L,
\end{equation}

and

\begin{equation}
\mbox{}^{L}u_{k}=0 \qquad\qquad   in R,
\end{equation}

\begin{equation}
\mbox{}^{L}u_{k}=(4\pi\nu)^{-1/2}e^{i(k\xi+\nu\eta)}\qquad\qquad  in L.
\end{equation}
   
Defining $\mbox{}^{L,R}v_{\nu}(\xi)=(4\pi\nu)^{-1/2}e^{ik\xi}$, it 
is possible to write the field operator as

\begin{eqnarray}
\varphi(\eta,\xi)=\int^{\infty}_{0}d\nu
&[&b^{(1)}(\nu)\mbox{}^{L}v(\nu,\xi)e^{i\nu\eta}+
b^{(1)^{\dagger}}(\nu)\mbox{}^{L}v^{\ast}(\nu,\xi)e^{-i\nu\eta}\nonumber\\
&+& b^{(2)}(\nu)\mbox{}^{R}v(\nu,\xi)e^{-i\nu\eta}+
b^{(2)^{\dagger}}(\nu)\mbox{}^{R}v(\nu,\xi)e^{i\nu\eta}].
\end{eqnarray}

Let us suppose that in $\eta_{i}=cte$ the state of the 
system is $\left|{\cal T}_{i}\right>=\left|g\right>
\otimes\left|0,M\right>$. The probability 
amplitude for the system to go to $\left|{\cal T}_{f}\right>
=\left|e\right>\otimes\left|\Phi_{f}\right>$ 
is:

\begin{eqnarray}
A^{1}_{|g>\otimes|0,M>\rightarrow|e>\otimes|\Phi_{f}>}&=&
-i\lambda_{1}m_{eg}(\eta_{i})\int^{\eta_{f}}_{\eta{i}}
d\eta\int^{\infty}_{0}d\nu[ \nonumber\\
e^{i(E-\nu)\eta}
(\mbox{}^{L}v^{\ast}(\nu,\xi)\left<\Phi_{f}\right|b^{(1)^{\dagger}}(\nu)
\left|O,M\right>&+&\mbox{}^{R}v(\nu,\xi)\left<\Phi_{f}
\right|b^{(2)}(\nu)
\left|O,M\right>)\nonumber\\
+ e^{i(E+\nu)\eta}(\mbox{}^{L}v(\nu,\xi)
\left<\Phi_{f}\right|b^{(1)}(\nu)\left|O,M\right>&+
&\mbox{}^{R}v(\nu,\xi)
\left<\Phi_{f}\right|b^{(2)^{\dagger}}(\nu)\left|O,M\right>)].
\end{eqnarray}

In the amplitude we have two terms. The first is an absorption 
process in the right edge and an
emission in the left edge of the Rindler manifold.
The second is an absorption 
process in the left edge and an
emission in the right edge of the Rindler manifold.
Nevertheless since this second one 
contains the factor $e^{i(E+\nu)\eta}$ which 
rapidly oscillates, it gives a negligible 
contribution to the amplitude for $\tau>>\frac{1}{|E|}$.

\section{Acknowledgement}

We would like to thank Prof.L.Ford, for valuable comments. This 
paper was supported by Conselho Nacional de Desenvolvimento 
Cient\'{\i}fico e
Tecnol\'ogico (CNPq) of Brazil.


\begin{thebibliography}{30}

\bibitem{Unruh} P.C.W.Davies, J.Phys.A {\bf 8}, 609 (1975), 
W.G.Unruh, Phys.Rev.D {\bf 14}, 870 (1976).
\bibitem{Nami1} B.F.Svaiter and N.F.Svaiter, 
Phys.Rev.D {\bf 46}, 5267 (1992),
ibid, Phys. Rev.D {\bf 47},4802 (1993)E.
\bibitem{DeWitt} B.DeWitt, in General Relativity - An Einstein Centenary
 Survey, edited by S.W.Hawking and W.Israel (Cambridge University Press, 
Cambridge, Englad, 1980).
\bibitem{Nami2} L.H.Ford, N.F.Svaiter and M.L.Lyra , 
Phys.Rev.A {\bf 49}, 178 (1994).
\bibitem{Nami3} B.F.Svaiter and N.F.Svaiter, Class.Quant.Grav. {\bf 11},
347 (1994).  
\bibitem{Ford} L.H.Ford and T.Roman, Phys.Rev.D {\bf 48}, 776 (1993).
\bibitem{Hinton} K.J.Hinton, J.Physics A {\bf 16}, 1937 (1983).
\bibitem{Kol} H.Kolbenstvedt, Phys.Rev.D {\bf 38}, 1118 (1988),
P.G.Grove, Class.Quant.Grav {\bf 3}, 801 (1986).
\bibitem{Ginzburg} V.L.Ginzburg and V.P.Frolov, Sov.Phys.
Usp. {\bf 30}, 1073 (1987).
\bibitem{Takagi} S.Takagi, Prog.Theor.Phys. {\bf 88}, 1 (1988).
\bibitem{Lebedev} N.N.Lebedev, "Special functions and their applications",
Dover Publication Inc, NY (1972).
\bibitem{Fulling} S.A.Fulling, Phys.Rev.D {\bf 7}, 2850 (1973).
\bibitem{Sciama} D.W.Sciama, P.Candelas and D.Deutsch, Adv. on 
Phys.{\bf 30}, 327 (1981). 
\bibitem{Allen} P.L.Knight and L.Allen,"Concepts of Quantum Optics" 
Pergamon Press Inc, USA (1985).
\bibitem{Peres} O.Levin, Y.Peleg and A.Peres, 
J.Phys.A {\bf 25}, 6471 (1992).
\bibitem{Unruh2} W.G.Unruh and R.Wald, Phys.Rev.D {\bf 29}, 1047 (1984).
\bibitem{Pringle} L.N.Pringle, Phys.Rev.D {\bf 39}, 2178 (1989),
T.D.Lee Nucl.Phys.{\bf B264}, 437 (1986).
\bibitem{Wilkens} M.Wilkens and M.Lewenstein, 
Phys.Rev.A, {\bf 39}, 4291 (1989).
\bibitem{Higuchi} A.Higuchi, G.E.A.Matsas and C.B.Peres,
Phys.Rev.D {\bf 48}, 3731 (1993).
\bibitem{Sri} L.Sriramkumar and T.Padmananabhan, Class.Quant.Grav.
{\bf 13}, 2061 (1996).
\bibitem{Landau} L.D.Landau and Peierls, selected
papers in Quantum Theory and Measurement, edited by J.A.Wheeler and 
W.H.Zurek, Princeton University Press N.Y. (1983).
\bibitem{Glauber} R.J.Glauber, Phys.Rev {\bf 130}, 2529 (1963),
ibid {\bf 131}, 2766 (1963).
\bibitem{Milonni} P.W.Milonni and W.A.Smith, Phys.Rev.A {\bf 11}, 814,
(1975).
\bibitem{Knight} J.R.Ackerhalt, P.L.Knight and J.H.
Eberly, Phys.Rev.Lett. {\bf 30}, 456 (1973).
\bibitem{Dicke} R.H.Dicke, Phys.Rev. {\bf 93}, 99 (1954).
\bibitem{Cohen} J.Dalibard, J.Dupont-Roc and C.Cohen-Tannoudji, 
J.Phys.(Paris) {\bf 43}, 1617 (1982); 
ibid J.Phys(Paris) {\bf 45}, 637 (1984).
\bibitem{Audretsch 1} J.Audretsch and R.Muller, 
Phys.Rev.A {\bf 50}, 1775 (1994),  J.Audretsch, 
R.Muller and M.Holzmann, 
Class.Quant.Grav. {\bf 12}, 2927 (1995),
J.Audretsch, M. Mensky and R.Muller, 
Phys.Rev.D {\bf 51}, 1716 (1995).
\bibitem{Barut} A.O.Barut and J.P.Dowling,
Phys.Rev.A {\bf 41}, 2277 (1990).
\bibitem{Pad} T.Padmanabhan and T.P.Singh, Class.Quant.Grav. {\bf 4}, 1397
(1987), T.Padmanabhan, Class.Quant. Grav. {\bf 2}, 117 (1985).
\bibitem{Candelas} P.Candelas and D.W.Sciama, 
Phys.Rev.D {\bf 27}, 1715 (1983). 
\bibitem{Letaw} J.R.Letaw and J.D.Pfautsch, Phys.Rev.D {\bf 22},
1345 (1980), ibid Phys.Rev.D, {\bf 24}, 1491 (1981), 
P.g.Grove and A.C.Otewill, J.Phys.A {\bf 16}, 3905 (1983), 
C.A.Manogue, Phys.Rev.D {\bf 35}, 3783 (1987).
\bibitem{Davies} P.C.W.Davies, T.Dray and C.A.Manogue, 
Phys.Rev.D {\bf 53}, 4382 (1996).
\bibitem{Takeno} H.Takeno, Prog.Theor.Phys. {\bf 7}, 367 (1952).
\bibitem{Miller} E.G.Kalnins and W.Miller, J.Math.Phys. {\bf 15},
1025 (1974), ibid J.Math.Phys. {\bf 19}, 1233 (1978).
\bibitem{Costa} I.Costa, J.Math.Phys. {\bf 30}, 888 (1989).
\bibitem{di} A.di Sessa, J.Math.Phys. {\bf 15}, 1892 (1974).
\bibitem{Som} C.M.Sommerfield, Ann.Phys. {\bf 84}, 285 (1974).




\end{thebibliography}
\end{document}